\documentstyle[pra,aps]{revtex} 

\begin{document} 
\draft 
\title{Narrow resonances with excitation of finite bandwidth field} 

\author{Peng Zhou$^{1, }$\thanks{Email: peng@qo1.am.qub.ac.uk}, 
Mao-Fa Fang$^{2,3,4}$,  Qing-Ping Zhou$^{5}$,  and Gao-Xiang Li$^{2,6,7}$} 
\address{$^{1}$Department of Applied Mathematics and Theoretical Physics,  \\
The Queen's University of Belfast, Belfast BT7 1NN, UK.\\
$^{2}$CCAST (World Laboratory), P.O. Box 8730, Beijing 100080, China.\\ 
$^{3}$Department of Physics, Hunan Normal University, 
Changsha 410081, China.\\
$^{4}$Anhui Institute of Optics and Fine Mechanics, Academia Sinica, Hefei
230031, China.\\
$^{5}$Department of Mechanics and Electronics, Wuling University, 
Zhangjiajie 427000, China. \\
$^{6}$Department of Physics and Astronomy, 
Vrije Universiteit, 1081 HV Amsterdam, Netherlands.\\
$^{7}$Department of Physics, Huazhong Normal University, Wuhan 430079, China.} 
\date{}
\maketitle 
 
\begin{abstract} 
The effect of the laser linewidth on the resonance fluorescence spectrum of a 
two-level atom is revisited. The novel spectral features, such as hole-burning 
and dispersive profiles at line centre of the fluorescence spectrum are 
predicted  when the laser linewidth is much greater than its intensity. 
The unique features result from quantum interference between different 
dressed-state transition channels. 
\end{abstract} 


The study of resonance fluorescence spectrum has provided much fundamental 
insight into the subject of atom-light interactions. It is well known that 
for weak laser field excitation, the spectrum exhibits a Lorentzian lineshape, 
whereas it develops into the Mollow triplet for strong field excitation 
\cite{mollow}.  The latter is a direct signature of stimulated emissions 
and absorptions of the atom during the time interval for one spontaneous 
decay.  

Recently, considerable attention has been paid to modifying the standard 
resonance fluorescence spectrum. Indeed, there are many ways to achieve this. 
One method is to place the atom inside a cavity, a wide variety of spectral 
features, such as dynamical suppression and enhancement, and spectral line 
narrowing of the Mollow triplet, has been predicted \cite{dynam} and detected 
\cite{obs}. Another method is to bathe the atom in a squeezed vacuum. 
Swain and co-worker \cite{ARF} then predicted anomalous fluorescence spectral 
features for weak excitation: hole-burning and dispersive profiles at line 
center, which are {\it qualitatively} different from any seen previously in
resonance fluorescence. Of late, Gawlik {\it et al.} \cite{coll} have shown
that rapidly elastic collisions between monochromatically driven atoms can 
also give rise to these anomalous profiles in resonance fluorescence.  

In this Letter we report that the anomalous fluorescence spectral features, 
such as hole-burning and dispersive profiles at line center, can even take 
place in a system of which a two-level atom is damped by a standard vacuum, 
and driven by a laser field with a finite bandwidth due to phase diffusions. 
The effect of the linewidth of the driving laser on the spectrum \cite{kimble} 
and the intensity fluctuations \onlinecite{wz,diode} of the resonance 
fluorescence had been extensively investigated both theoretically and 
experimentally. However, most of these studies concentrated on the case of 
which the laser intensity is much greater than its linewidth, where the 
spectral components are well resolved. The spectral line broadening, 
suppression and asymmetry in the Mollow triplet were reported \cite{kimble}. 
We are here mainly interested in the region of which the laser linewidth is 
larger than its intensity.  Novel resonance lineshapes, --- the hole-burning 
and dispersive profiles at the spectral line  centre of the resonance 
fluorescence are predicted. 

We consider a single two-level atom with transition frequency $\omega_{A}$ 
driven by a laser field with amplitude ${\cal E}$ and frequency 
$\omega_{L}$ and fluctuating phase $\phi(t)$. The master equation for 
the atomic density matrix operator $\rho$, in a frame rotating at the 
frequency $\omega_{L}$, is of the form 

\begin{equation} 
\dot{\rho}=-i\left[ H_{AL},\,\rho \right] +{\cal L}\rho,  
\end{equation} 
where 
\begin{eqnarray} 
H_{AL} &=& \frac{\Delta }{2}\sigma_{z}+\frac{\Omega }{2}
\left[ e^{-i\phi(t)}\sigma_{+}+e^{i\phi(t)}\sigma_{-}\right], \\ 
{\cal L}\rho  &=& \gamma (2\sigma_{-}\rho \sigma_{+}-\sigma_{+}\sigma_{-}
\rho -\rho \sigma_{+}\sigma_{-}),  
\end{eqnarray} 
where $H_{AL}$ is the Hamiltonians of the coherently driven atom and 
${\cal L}\rho $ describes the atomic spontaneous decay with the rate 
$\gamma $,  $\sigma_{\pm}$ and $\sigma_z$ are the atomic upper (lower) 
transition and population inversion operators, respectively, 
$\Omega =2|\mu_{01}{\cal E}|/\hbar$ is the driving Rabi frequency, 
$\Delta=\omega_{A}-\omega_L$ is the detuning between the atomic transition 
and the driving laser. The fluctuating phase,  $\phi(t)$,  results in a 
stochastic frequency $\vartheta(t)=\dot{\phi}(t)$, which is assumed to be a 
Gaussian random process with the properties \onlinecite{kimble,diode}

\begin{equation} 
\langle \vartheta(t) \rangle=0, \hspace{0.5cm}
\langle \vartheta(t) \vartheta(t') \rangle= 2L \delta (t-t'), 
\end{equation} 
where $L$ is the strength of the frequency fluctuations and physically
describes the effective bandwidth of the laser beam due to the phase
diffusion. This is the situation most appropriate for describing the radiation from a diode
laser, which has a very stable amplitude and very large phase-diffusions when
it is operated far above threshold \cite{diode}.  After averaging over the 
stochastic phase, one can obtain the optical Bloch equation to be 
\cite{kimble} 

\begin{eqnarray} 
&&\langle \dot{\sigma}_{-}\rangle =-(\Gamma +i\Delta) \langle \sigma_{-}\rangle 
  +\frac{i}{2}\Omega \langle \sigma_z\rangle , \nonumber\\ 
&& \langle \dot{\sigma}_z \rangle =- \gamma_z\langle \sigma_z \rangle + i
\Omega\left( \langle \sigma_{-} \rangle- \langle \sigma_{+} \rangle\right)
-\gamma_z ,  \label{bloch} 
\end{eqnarray} 
where $\Gamma=\gamma + L$ and $\gamma_z =2\gamma$  represent respectively the
transverse and longitudinal relaxation rates.

The resonance fluorescence spectrum in the far radiation zone 
can be expressed, in term of the steady-state atomic correlation function, 
as \cite{kimble} 
\begin{equation}
\Lambda(\omega) =\mbox{Re}  \int_0^\infty \lim_{t\rightarrow \infty} 
\langle \sigma_{+}(t+\tau)  \sigma_{-}(t) 
\rangle e^{-i\omega \tau} d\tau  =\mbox{Re} [{\cal D}(i\omega)],
\label{rfs1}
\end{equation}
where ${\cal D}(z)$ is the Laplace transform of the atomic correlation function
$\lim_{t\rightarrow \infty} \langle \sigma_{+}(t+\tau)  \sigma_{-}(t)
\rangle$, which is obtained, by invoking the quantum regression theorem,
together with the Bloch equations (\ref{bloch}), to be 

\begin{equation} 
{\cal D}(z)=\frac{\left[\left(\Gamma +i\Delta +z\right) (\gamma_z +z) 
+\Omega^2/2\right] (1+\langle \sigma_{z}\rangle_{s}) 
+i\Omega \left(\Gamma +i\Delta+z\right) (1+\gamma_z/z) 
\langle\sigma_{-}\rangle_{s} }
{2(\gamma_z +z)\left[(\Gamma+z)^2 + \Delta^2\right] +2\Omega^2(\Gamma+z)}, 
\label{rfs2}
\end{equation}
where $\langle\sigma_{-}\rangle_{s}$ and $\langle\sigma_{z}\rangle_{s}$
are the steady-state solutions of the Bloch equation (\ref{bloch}).

Our formulae (\ref{rfs1})-(\ref{rfs2}) can reproduce the previous predictions 
of Mollow \cite{mollow} (when $L=0$) and Kimble {\it et al.} \cite{kimble} 
(when $\Omega \gg L$). However, here we exploit novel spectral features 
in the regime of $L \geq \Omega$, which has been paid little attention in 
the past.  

Figure 1 presents the resonance fluorescence spectrum of the atom with
excitation of a resonant ($\Delta=0$), strong laser field ($\Omega=50\gamma \gg 
\gamma$) with various laser linewidths ($L=10\gamma,\,50\gamma,\,100\gamma,\,
200\gamma$). It is obvious that when the laser linewidth $L$ is much less than
the Rabi frequency $\Omega$, see, for example, in the frame (a) for 
$L=10\gamma$, as Kimble {\it et al.} \cite{kimble} predicted, the spectrum 
still exhibits a three-peak structure, but with a suppressed central peak and 
narrowed sidebands, comparing to the standard Mollow triplet \cite{mollow}.   
As the laser linewidth widens, the central peak is greatly suppressed while the
sideband resonances are merged, therefore, a dip ({\it i.e.}, a hole burning 
profile) places at line centre, ---see in Fig. 1(b) where $L=50\gamma$. 
When the value of the laser linewidth is much larger than the Rabi frequency, 
the dip becomes very narrow, as depicted in Figs. 1(c) and 1(d) where 
$L=100\gamma$ and $200\gamma$, respectively.  

Figure 2 illustrates a three dimensional fluorescence spectrum against the laser
linewidth, for $\gamma =1,\, \Omega=50\gamma$ and $\Delta=0$, from which one 
can see how the Mollow triplet is suppressed and the dip is developed at line
centre as the laser linewidth increases.   

Figure 3 clearly shows that the resonance fluorescence spectrum may exhibit 
another narrow resonance feature---dispersive (Rayleigh-wing) profile, when 
the laser with a very wide linewidth is appropriately detuned from the atomic 
transition frequency. We have taken the parameters $\gamma=1,\, 
\Omega=50\gamma,\, L=200\gamma$ in Fig. 3. When the laser-atom detuning is 
comparable with the laser linewidth, {\it e.g.}, in Figs. 3(b)-3(c) for 
$\Delta=100\gamma,\,200\gamma$, the dispersive profile is most pronounced, 
otherwise,  it is less pronounced, see, for instance, in Fig. 3(a) 
($\Delta =50\gamma \ll L $)  and Fig. 3(d) ($\Delta =400\gamma \gg L $). 
The latter frame demonstrates the narrow peak at the laser frequency (line 
centre) and a broad peak at the atomic transition frequency. The both 
resonances are well split, which is the case of Kimble {\it et al.} 
\cite{kimble}.

When the laser bandwidth is much larger than the other parameters, 
{\em i.e.} $\Gamma \gg \Omega,\,\Delta, \, \gamma_z$,  the
resonance fluorescence spectrum (\ref{rfs1}) approximately takes the form, 

\begin{equation}
\Lambda (\omega)\simeq \frac{\Gamma}{4(\gamma_z\Gamma +\Omega^2)}
\left[\frac{\Omega^2-2\Delta\omega}{\Gamma^2+\omega^2} 
+\frac{\Omega^2+2\Delta\omega}{\left(\Gamma-\Omega^2/\Gamma\right)^2+\omega^2}
-\left(\frac{\Omega}{\Gamma} \right)^4 \frac{\Omega^2+2\Delta\omega}
{\left(\gamma_z+\Omega^2/\Gamma\right)^2+\omega^2}\right],
\label{app}
\end{equation}
which  consists of three resonances located at line
center, but with different resonance linewidths. The first two resonances have 
positive weights and linewidths of an order of $2\Gamma$ (noting that 
$\Omega^2/\Gamma \ll 1$), whereas the last one has a very narrow linewidth of
$2\gamma_z$, comparing to $2\Gamma$,  and a negative weight. The latter gives
rise to  a typical spectral profile of which a narrow hole is bored into a 
broad peak. The approximate expression (\ref{app}) of the resonance 
fluorescence spectrum also shows that when the laser is resonant with the atom
($\Delta=0$), all the three resonances are the Lorentzian lineshape, therefore,
the spectrum is symmetry, as shown in Figs. 1-2. Otherwise, these resonances mix
up the Lorentzian and Rayleigh-wing (dispersive) lineshapes when the laser is
detuned from the atom. As a results, the spectrum exhibits asymmetry, see, for
example in Fig. 3. 

The hole-burning and dispersive profiles at the spectral centre of the 
resonance fluorescence are attributed to quantum interference \cite{inter}. 
To explain this, we work in the basis of the semiclassical dressed states 
$|\pm \rangle $, defined by the eigenvalue equation $H_{AL} |\pm \rangle =
\pm (\bar{\Omega}/2)|\pm\rangle $. For simplicity, we consider ony the case 
of $\Delta=0$. The dressed states are then given by  $|\pm \rangle = 
(|0\rangle \pm |1\rangle)/\sqrt{2}$. In the limit of $\Omega \gg \gamma $, 
the equations of motion then simplify to

\begin{eqnarray}
&& \langle \dot{R}_{++} \rangle=-\Gamma \langle R_{++} \rangle +\frac{\Gamma}
{2},\label{dr1}\\
&& \langle \dot{R}_{--} \rangle=-\Gamma \langle R_{--} \rangle +\frac{\Gamma}
{2},\label{dr2}\\
&& \langle \dot{R}_{+-} \rangle =-\left(\Gamma_{+} -i\Omega \right) 
\langle R_{+-} \rangle +\Gamma_{-}\langle  R_{-+} \rangle, \label{dr3}\\
&& \langle \dot{R}_{-+} \rangle =-\left(\Gamma_{+} +i\Omega \right) 
\langle R_{-+} \rangle +\Gamma_{-}\langle  R_{+-} \rangle, \label{dr4}
\end{eqnarray}
where $\Gamma_{\pm} =(\Gamma \pm \gamma_z)/2$, $R_{lk} =|l\rangle \langle 
k|$ ($l,k=\pm$) is an atomic downward transition operator between the dressed 
states $|l\rangle $ and $|k\rangle $ of two near-lying dressed doublets.  
Eqs. (\ref{dr1}) and (\ref{dr2}) describe the atomic downward transitions 
between the same dressed states of two adjacent dressed doublets, whereas, 
eq. (\ref{dr3}) (eq. (\ref{dr4})) represents transitions from the dressed 
state $|+\rangle$ ($|-\rangle$) of one dressed doublets to the dressed state 
$|-\rangle$ ($|+\rangle$) of the next dressed doublets. 
If the Rabi frequency $\Omega$ is much greater than $\Gamma_{\pm}$, the terms
associated with different resonant frequencies, 
$\Gamma_{-}\langle  R_{-+} \rangle$ in eq. (\ref{dr3}) and 
$\Gamma_{-}\langle  R_{+-} \rangle$ in eq. (\ref{dr4}) are negligible  
under the secular approximation \cite{cohen}. Consequently, the both 
transitions, $|+\rangle \rightarrow |-\rangle $ and $|-\rangle \rightarrow 
|+\rangle $ are independent. Otherwise, they are correlated \cite{cor}, 
{\it i.e.}, as the atom decays from $|+\rangle$ to $ |-\rangle $ it drives 
the other transition from $|-\rangle$ to $ |+\rangle$, and vice versa. 
This reflects the fact that fluorescent photons emitting from these 
transitions are indistinguishable so that quantum interference between these 
transition channels \cite{inter} dominates.

It is well known that the resonance fluorescence can be described by 
spontaneous emissions of the atom downward the ladder of the dressed-state 
doublet \cite{cohen}. The atomic decays between same dressed states of
two adjacent dressed doublets, governed by eqs. (\ref{dr1}) and (\ref{dr2}),
give rise to a spectral component  

\begin{equation}
\Lambda_0(\omega) =\frac{\Gamma}{4\left(\Gamma^2+\omega^2\right)},
\end{equation}
which centres at the laser frequency and has a linewidth $2\Gamma$ and a 
height $1/(4\Gamma)$. The spectrum broadens and is suppressed as the laser
linewidth increases. 

Whereas, the other transitions, described by eqs. (\ref{dr3}) and (\ref{dr4}), 
result in a spectrum 

\begin{equation}
\Lambda_{1}(\omega) =\frac{1}{4} \mbox{Re}\left[ \frac{\gamma_z+i\omega}
{(\Gamma_{+}+i\omega)^2+\Omega^2-\Gamma_{-}^2} \right],
\end{equation}
whose position and feature are dependent on the laser linewidth $L$ and
intensity $\Omega$. 

The total fluorescence spectrum consists of the two spectra, $\Lambda(\omega)=
\Lambda_0(\omega)+\Lambda_1(\omega)$, which are demonstrated in Fig. 4 
for different laser linewidths. The spectrum $\Lambda_0(\omega)$ always shows a
Lorentzian shape located at line centre, which is independent of the laser
intensity, but varies with the laser linewidth. As the linewidth increases the
spectral height is suppressed and the spectral width is brodened. Whereas,  
$\Lambda_1(\omega)$ is sensitive to the both parameters. 
When $L \ll \Omega$, $\Lambda_1(\omega)$ exhibits a well-resolved, two-peak 
structure, as shown in Fig. 4a. This is because the dressed doublet $|\pm
\rangle $ is well separated, and the resultant transitions $|+\rangle 
\rightarrow |-\rangle $ and $|-\rangle \rightarrow |+\rangle $ have different
(distinguished) resonance frequencies $\omega \pm \Omega$. Correspondently, the
total fluorescence spectrum has a three-peak Mollow structure. When $L \geq
\Omega$, $\Lambda_1(\omega)$ shows a dip bored into 
a wide bell-shape spectrum \cite{agr}, as depicted in Figs. 4b--4d. The total
spectrum thus has a hole burning profile at line centre.  The larger the laser 
linewidth is, the narrower the hole burning is. It is obvious that the 
hole burning (dip) profile originates from the correlated transitions (quantum 
interference) between the dressed states $|\pm \rangle $ in two adjacent 
dressed doublets. 

When the atom-laser detuning is taken into account, the populations in the
dressed states  $|\pm \rangle $ are not same. Hence, the transitions  
$|+\rangle \rightarrow |-\rangle $ and $|-\rangle \rightarrow |+\rangle $ have
different probability amplitudes. The resultant fluorescence spectrum is
asymmetric. The total spectrum thus exhibits a dispersive-like profile at line
centre.  

In summary, we demonstrate that when a two-level atom is excited with a strong
laser field with a broad bandwidth due to phase diffusions, the resonance 
fluorescence spectrum may exhibit the anomalous, narrow resonance features,  
such as hole-burning and dispersive profiles at line centre. The physics behind
the anomalous spectral features is quantum interference between different
dressed-state transition channels. From the experimental point of view, 
observing  these features in the system is much easier than that in a squeezed 
vacuum \cite{ARF}.

\acknowledgements  
 
This work is supported by the National Natural Science Foundation of China and
the United Kingdom EPSRC. P.Z. wishes to thank S. Swain for discussions 
and W. Gawlik for providing with his reprint.

\begin{figure}[tbp] 
\caption{The resonance fluorescence spectrum $\Lambda(\omega)$ for $\gamma=1,\,
\Omega=50\gamma,\, \Delta=0$, and different laser bandwidths: $L=10\gamma$ (a), $L=50\gamma$ (b), 
$L=100\gamma$ (c), $L=200\gamma$ (d).} 
\label{fig1} 
\end{figure} 

\begin{figure}[tbp] 
\caption{Three dimensional fluorescence spectrum $\Lambda(\omega)$ for $\gamma=1,
\,\Omega=50\gamma,\, \Delta=0$. } 
\label{fig2} 
\end{figure}

\begin{figure}[tbp] 
\caption{Same as FIG. 1, but for $\gamma=1,\,\Omega=50\gamma,\, 
L=200\gamma$, and different laser-atom detunings: $\Delta=50\gamma$ (a), 
$\Delta=100\gamma$ (b), $\Delta=200\gamma$ (c), $\Delta=400\gamma$ (d).} 
\label{fig3} 
\end{figure} 

\begin{figure}[tbp] 
\caption{Spectral components, $\Lambda_0(\omega)$ (dotted curve) and 
$\Lambda_1(\omega)$ (dashed curve), and
the total fluorescence spectrum $\Lambda(\omega)= \Lambda_0(\omega)+
\Lambda_1(\omega)$  (solid curve), where values of the parameters are set same 
as those in Fig. 1.} 
\label{fig4} 
\end{figure} 


\begin{references} 
\bibitem{mollow}  
B. R. Mollow, Phys. Rev. {\bf 188} (1969) 1969;\\
F. Y. Wu, R. E. Grove and S. Ezekiel, Phys. Rev. Lett. {\bf 35} (1975) 1426.

\bibitem{dynam}  
M. Lewenstein, T. W. Mossberg and R. J. Glauber, Phys. Rev. Lett. 
{\bf 59} (1987) 775; \\
P. Zhou and S. Swain, Phys. Rev. A {\bf 58} (1998) 1515. 

\bibitem{obs}
Y. Zhu, A. Lezama and T. W. Mossberg, Phys. Rev. Lett. {\bf 61} (1988) 1946.  

 
\bibitem{ARF}  
S. Swain, Phys. Rev. Lett. {\bf 73} (1994) 1493;\\ 
S. Swain and P. Zhou, Phys. Rev. A {\bf 52} (1995) 4845;\\ 
P. Zhou, Z. Ficek and S. Swain, J. Opt. Soc. Am. B {\bf 13} (1996) 768. 
 
\bibitem{coll}
W. Gawlik, B. Lobodzinski and W. Chalupczak, in:
Fronters of Quantum Optics and Laser Physics, eds. S. Y. Zhu, 
M. O. Scully and M. S. Zubairy, (Springer, 1997).
       

\bibitem{kimble}
G. S. Agarwal, Phys. Rev. Lett. {\bf 37} (1976) 1383;\\
J. H. Eberly, Phys. Rev. Lett.  {\bf 37} (1976) 1387;\\
H. J. Kimble and L. Mandel, Phys. Rev. A {\bf 15} (1977) 689;\\ 
P. L. Knight,  W. A. Molander and C. R. Stroud, Jr.,  Phys. Rev. A 
{\bf 17} (1978) 1547;\\
S. Swain, Adv. At. Mol. Opt. {\bf 16} (1980) 159; \\
A. H. Toor and M. S. Zubairy,  Phys. Rev. A {\bf 49} (1994) 449.

\bibitem{wz}
M. H. Anderson, R. D. Jones, J. Cooper, S. J. Smith, D. S. Elliot, H. Ritsch 
and P. Zoller,  Phys. Rev. Lett. {\bf 64} (1990) 1346; 
 Phys. Rev. A {\bf 42} (1990) 6690;\\
R. Walser and P. Zoller, Phys. Rev. A {\bf 49} (1994) 5067.

\bibitem{diode}
T. Yabuzaki, T. Mitsui and U. Tanaka, Phys. Rev. Lett. {\bf 67} (1991) 2453;\\
K. V. Vasavada,  G. Vemuri and G. S. Agarwal, Phys. Rev. A {\bf 52} (1995) 4159.

\bibitem{cohen}  
C. Cohen-Tannoudji and  S. Reynaud, J. Phys. B {\bf 10} (1977) 345.

\bibitem{inter}
S. Y. Zhu, L. M. Narducci and M. O. Scully, Phys. Rev. A  {\bf 52} (1995) 4791;\\ 
P. Zhou and S. Swain, Phys. Rev. Lett. {\bf 78} (1997) 832; 
 J. Opt. Soc. Am. B {\bf 15} 1998) 2593.

\bibitem{cor}
A. Aspect, G. Roger, S. Reynaud, J. Dalibard and C. Cohen-Tannoudji, 
Phys. Rev. Lett. {\bf 45} (1980) 617; \\
C. A. Schrama, G. Nienhuis, H. A. Dijkerman, C. Steijsiger and H. G. M. Heideman, 
Phys. Rev. Lett. {\bf 67} (1991) 2443.

\bibitem{agr}
G. S. Agarwal, in: Fronters of Quantum Optics and Laser Physics, 
eds. S. Y. Zhu, M. O. Scully and M. S. Zubairy, (Springer, 1997).

\end{references}
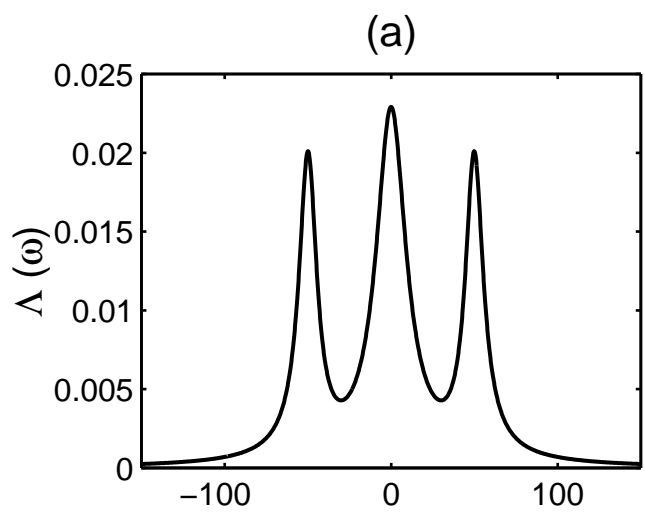
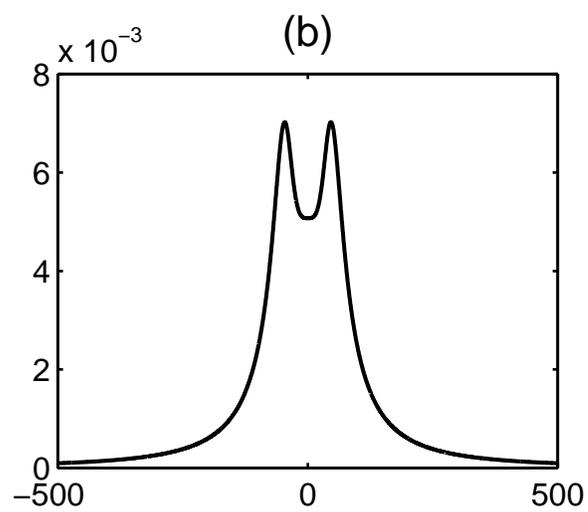
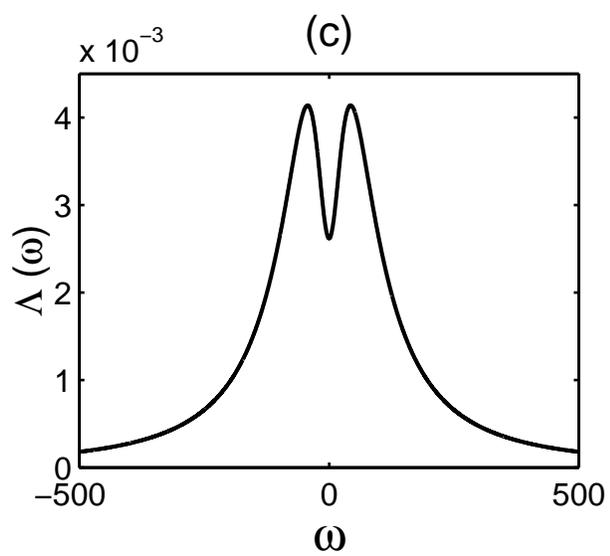
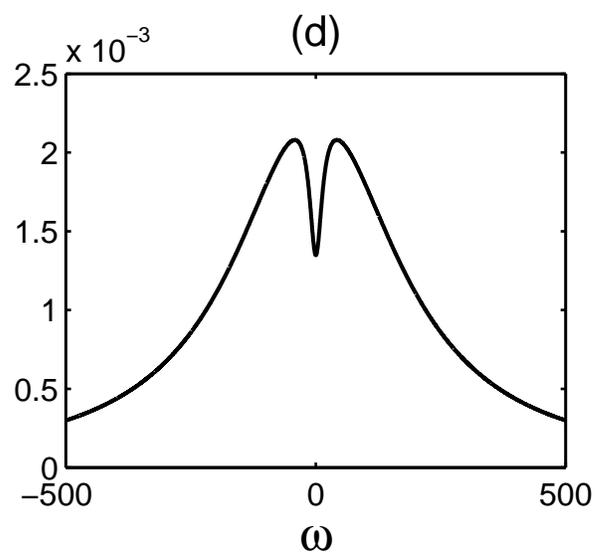
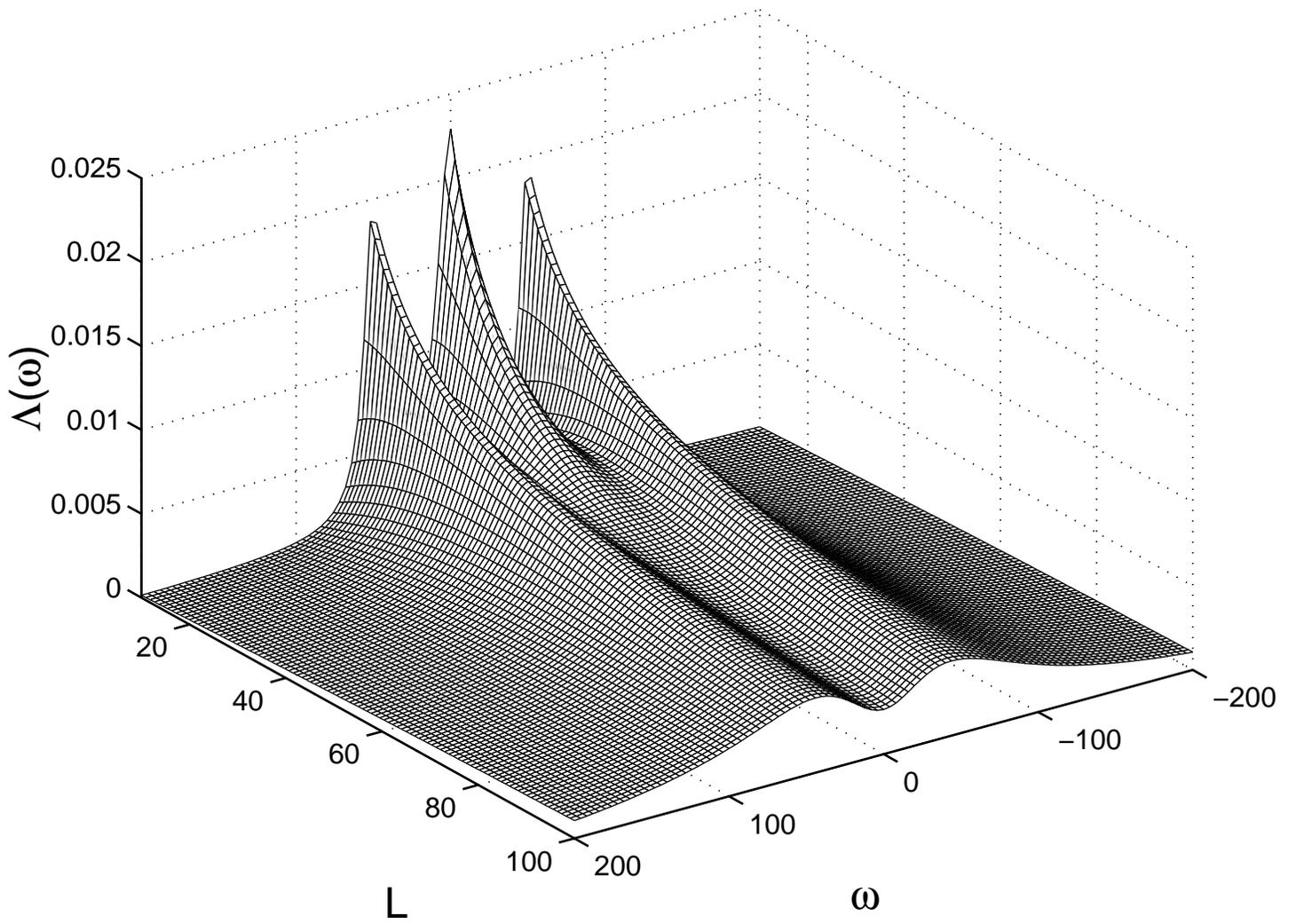
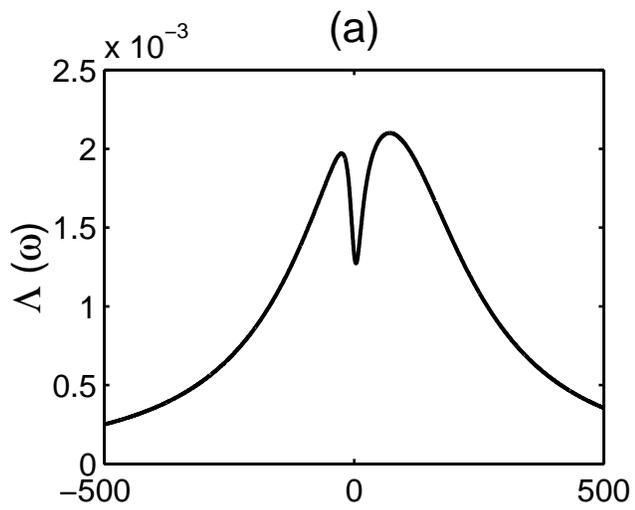
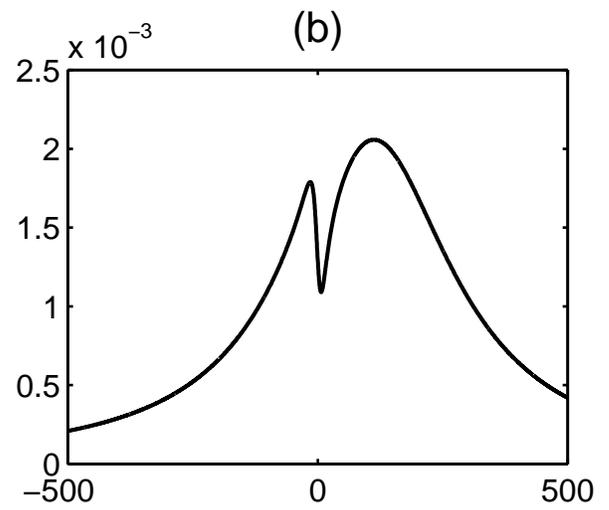
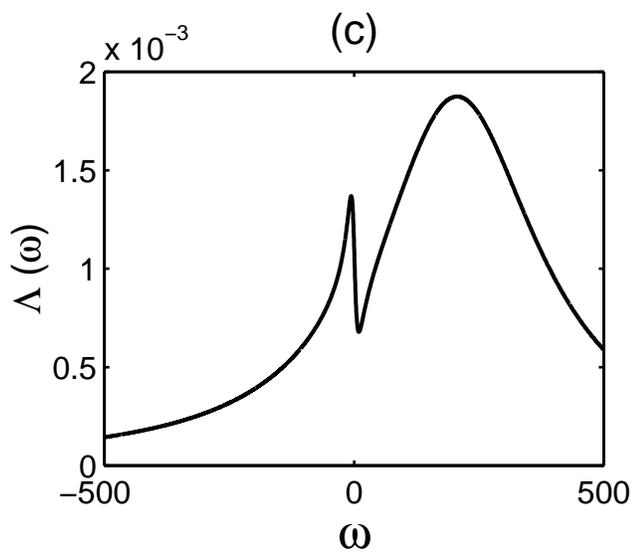
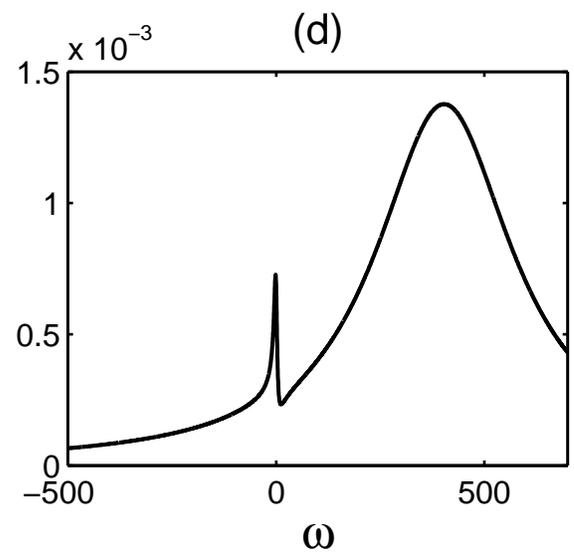
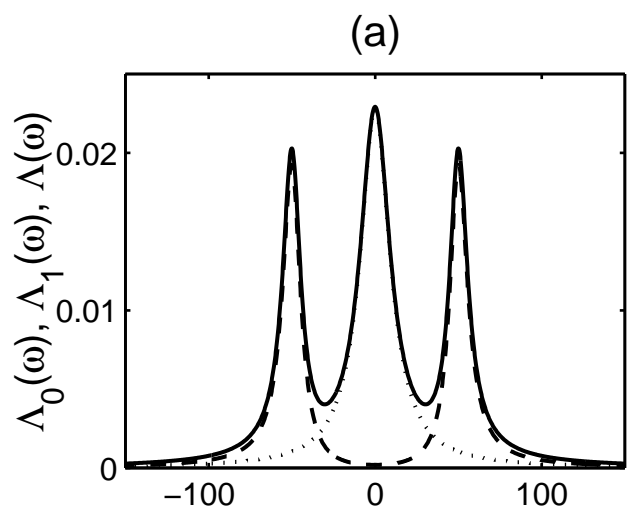
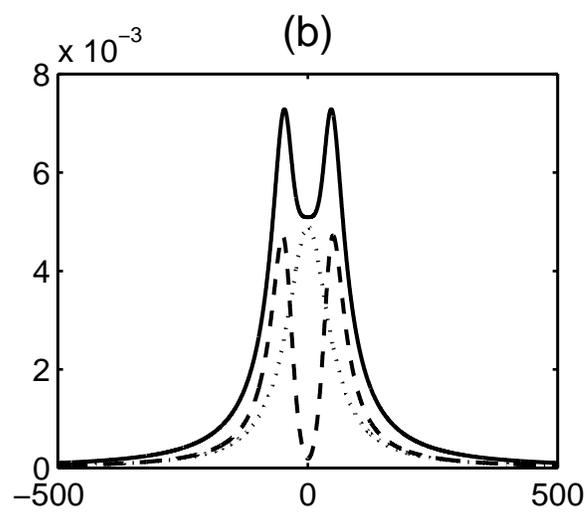
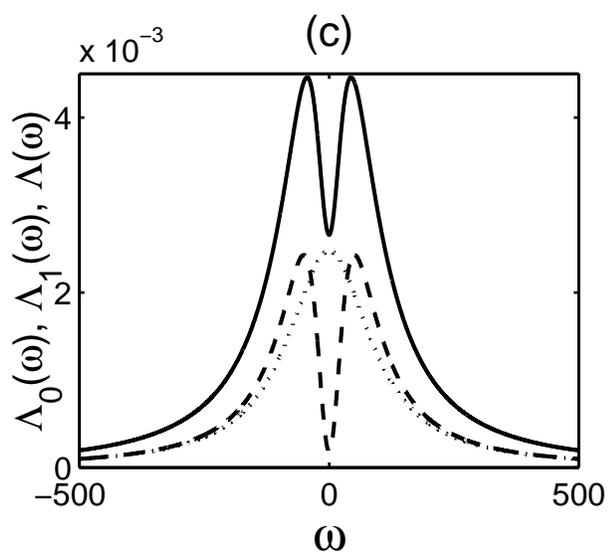
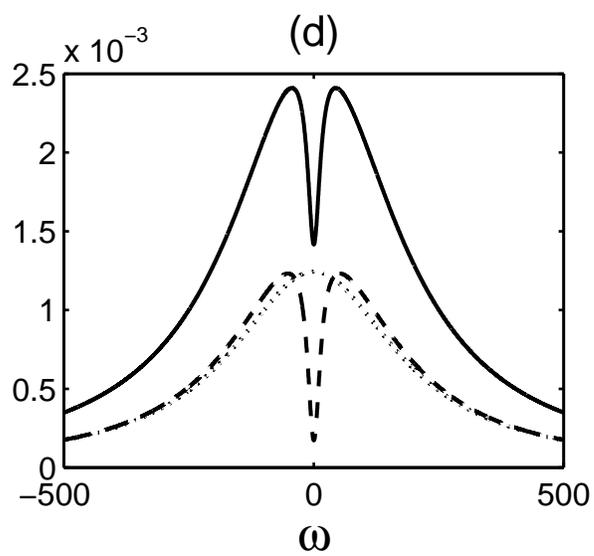
\end{document}